\documentclass[aps,prd,reprint,superscriptaddress]{revtex4-2}

\usepackage{amsmath} \usepackage{graphicx} \usepackage{subfigure}
\usepackage{epstopdf} \usepackage{color} \usepackage{multirow}
\usepackage{setspace} \usepackage{overpic} \usepackage{amssymb}
\usepackage{ulem}
\usepackage[bookmarksnumbered, pdfstartview=FitH,colorlinks,urlcolor=blue, citecolor=blue,linkcolor=blue] {hyperref}
\usepackage{lineno}
\usepackage{bm}
\usepackage{rotating}
\usepackage{xcolor}
\usepackage{makecell}
\usepackage{mathtext}
\usepackage{mathrsfs}
\usepackage[T1]{fontenc}
\usepackage{lmodern}
\usepackage[utf8]{inputenc}
\usepackage{besphysics}

\newcommand{\jnp}{J/\psi \to p \pi^{-} \bar{n}}
\newcommand{\BESIIIorcid}[1]{\href{https://orcid.org/#1}{\hspace*{0.1em}\raisebox{-0.45ex}{\includegraphics[width=1em]{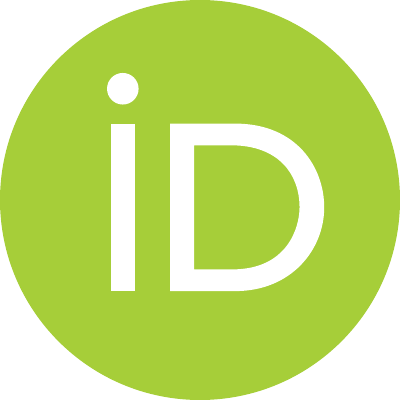}}}}

\begin{document}

\title{\texorpdfstring{\boldmath Cross Section Measurements of $\bar{n}p \rightarrow K^{+}K^{-}\pi^{+}(\pi^{0})$ via Antineutrons Produced by $\jnp$ Decays}{Cross Section Measurements of antineutron-proton to K+K-pi+(pi0) via Antineutrons Produced by J/psi to p pi- antineutron Decays}}
\author{
    \begin{center}
        %% Saved at => 2025-10-27
M.~Ablikim$^{1}$\BESIIIorcid{0000-0002-3935-619X},
M.~N.~Achasov$^{4,c}$\BESIIIorcid{0000-0002-9400-8622},
P.~Adlarson$^{81}$\BESIIIorcid{0000-0001-6280-3851},
X.~C.~Ai$^{87}$\BESIIIorcid{0000-0003-3856-2415},
C.~S.~Akondi$^{31A,31B}$\BESIIIorcid{0000-0001-6303-5217},
R.~Aliberti$^{39}$\BESIIIorcid{0000-0003-3500-4012},
A.~Amoroso$^{80A,80C}$\BESIIIorcid{0000-0002-3095-8610},
Q.~An$^{77,64,\dagger}$,
Y.~H.~An$^{87}$\BESIIIorcid{0009-0008-3419-0849},
Y.~Bai$^{62}$\BESIIIorcid{0000-0001-6593-5665}
% ...
\\
\vspace{0.2cm}
(BESIII Collaboration)\\
\vspace{0.2cm} {\it
$^{1}$ Institute of High Energy Physics, Beijing 100049, People's Republic of China\\
% ...
$^{4}$ Budker Institute of Nuclear Physics SB RAS (BINP), Novosibirsk 630090, Russia\\
% ...
$^{31}$ INFN Sezione di Ferrara, (A)INFN Sezione di Ferrara, I-44122, Ferrara, Italy; (B)University of Ferrara, I-44122, Ferrara, Italy\\
% ...
$^{39}$ Johannes Gutenberg University of Mainz, Johann-Joachim-Becher-Weg 45, D-55099 Mainz, Germany\\
% ...
$^{62}$ Southeast University, Nanjing 211100, People's Republic of China\\
% ...
$^{64}$ State Key Laboratory of Particle Detection and Electronics, Beijing 100049, Hefei 230026, People's Republic of China\\
% ...
$^{77}$ University of Science and Technology of China, Hefei 230026, People's Republic of China\\
% ...
$^{80}$ University of Turin and INFN, (A)University of Turin, I-10125, Turin, Italy; (B)University of Eastern Piedmont, I-15121, Alessandria, Italy; (C)INFN, I-10125, Turin, Italy\\
$^{81}$ Uppsala University, Box 516, SE-75120 Uppsala, Sweden\\
% ...
$^{87}$ Zhengzhou University, Zhengzhou 450001, People's Republic of China\\

\vspace{0.2cm}
$^{\dagger}$ Deceased\\
% ...
$^{c}$ Also at the Novosibirsk State University, Novosibirsk, 630090, Russia\\
% ...

}
%% ends here %%

    \end{center}
}

\date{\today}
% \linenumbers
\begin{abstract}
    Based on a novel method for producing antineutrons via $J/\psi$ decays, we report a study of $\bar{n}p$ inelastic scattering into final states containing kaons. The analysis uses $(10087\pm44)\times 10^6$ $J/\psi$ events collected at the BESIII detector operating at the BEPCII storage ring. Antineutrons are produced via $\jnp$ decays and tagged by the detected protons and pions, resulting in antineutron momenta ranging from 0 to 1174$\mevc$, while target protons are provided by the hydrogen in the beam-pipe material. The flux-averaged cross sections of the reactions $\bar{n}p \rightarrow K^{+}K^{-}\pi^{+}$ and $\bar{n}p \rightarrow K^{+}K^{-}\pi^{+}\pi^{0}$ over this antineutron momentum spectrum are measured to be $0.53^{+0.15}_{-0.12} \pm 0.08$~mb and $1.09^{+0.36}_{-0.30} \pm 0.31$~mb, respectively, where the first uncertainties are statistical and the second are systematic. Due to limited statistics, the intermediate states in these processes are not quantitatively investigated. The observation of clean antineutron-proton scattering candidates indicates the potential of this approach for future investigations of antineutron-proton interactions.
\end{abstract}
\maketitle

\section{Introduction}
Scattering experiments are powerful tools for probing the microstructure of matter and the interactions among particles.  They have been widely used since Rutherford aimed $\alpha$ particles at a gold target in 1911~\cite{E.Rutherford}. As fundamental constituents of matter, nucleons and their antiparticles, antinucleons, have attracted extensive research interest. In particular, the development of antiproton beams at LEAR made it possible to conduct comprehensive studies of antiproton-proton~($\bar{p}p$) scattering~\cite{Amsler:2019ytk}, including cross section measurements, meson spectroscopy studies, and searches for bound $\bar{p}p$ states, significantly advancing our understanding of nucleon-antinucleon interactions.

Unlike $\bar{p}p$ scattering, antineutron-proton ($\bar{n}p$) scattering involves only the $I = 1$ state and is free from significant Coulomb interactions. These advantages make it an ideal laboratory for studying the strong interactions of the nucleon and antinucleon. Although antiproton-neutron ($\bar{p}n$) scattering shares these advantages, neutron targets are experimentally challenging due to the presence of surrounding protons. Moreover, due to the substantial difficulties in producing and controlling antineutron beams, $\bar{n}p$ scattering experiments are much less than $\bar{p}p$ experiments. 
The traditional method for producing antineutron beams is via the charge exchange (CEX) reaction $\bar{p}p \rightarrow \bar{n}n$. This technique has been used in experiments such as BNL E-767 (with antineutron momentum between 100 and 500$\mevc$)~\cite{BNL} and CERN OBELIX (with antineutron momentum between 50 and 400$\mevc$)~\cite{OBELIX}. These antineutron beams have enabled important investigations, yielding intriguing results, particularly in strangeness production. For example, a notable violation of the OZI rule was observed in $\bar{n}p \rightarrow \phi \pi^+$ at antineutron momentum below 300$\mevc$~\cite{OBELIX:1994hbn}, where $\phi$ production greatly exceeded expectations. Additionally, a prominent di-kaon decay mode of the $f_0(1500)$ was observed; this state is a candidate for having a large scalar glueball component.

Although previous experiments have significantly advanced our understanding of antinucleon-nucleon interactions, antineutron scattering data remain limited because of the low production rate of antineutron via the CEX reaction. Recently, a novel method for producing antineutrons and hyperons has been proposed~\cite{Yuan,Dai:2024myk}, in which the $J/\psi$ meson serves as an efficient antineutron source. This approach takes advantage of the appreciable branching fractions of $J/\psi$ decays to final states containing antineutrons and the high production cross section of $J/\psi$ in $e^+e^-$ annihilation. Antineutrons produced through $\jnp$ decays cover a broad momentum range up to 1174$\mevc$, extending the range beyond previous experiments. Using this method, the BESIII Collaboration has measured the cross sections for $\bar{n}p$ inelastic scattering into pionic states~\cite{PWA}, and has also reported studies of hyperon and antihyperon scattering with nucleons and nuclei~\cite{BESIII:2023clq,BESIII:2023trh,BESIII:2024geh,BESIII:2025bft}.  

In this paper, we report measurements of the flux-averaged cross sections for the reactions $\bar{n}p \rightarrow K^{+}K^{-}\pi^{+}$ and $\bar{n}p \rightarrow K^{+}K^{-}\pi^{+}\pi^0$ using this novel method at the BESIII detector, based on a data sample of $(10087 \pm 44) \times 10^{6}$ $J/\psi$ events~\cite{jpsi data}. Antineutrons are produced via $\jnp$ with a branching fraction of 0.212(9)\%, and tagged by the recoil of the detected $p \pi^-$ particles, while the target protons originate from the hydrogen in the oil layer of the beam pipe. 

\section{BESIII detector and Monte Carlo simulation}
The BESIII detector~\cite{Ablikim:2009aa} records symmetric $e^+e^-$ collisions 
provided by the BEPCII storage ring~\cite{Yu:IPAC2016-TUYA01}
in the center-of-mass energy range from 1.84 to 4.95~GeV, with a peak luminosity of $1.1 \times 10^{33}\;\text{cm}^{-2}\text{s}^{-1}$ achieved at $\sqrt{s} = 3.773\;\text{GeV}$. BESIII has collected large data samples in this energy region~\cite{Ablikim:2019hff,EcmsMea,EventFilter}. The cylindrical core of the BESIII detector covers 93\% of the full solid angle and consists of a helium-based multilayer drift chamber~(MDC), a time-of-flight system~(TOF), and a CsI(Tl) electromagnetic calorimeter~(EMC), which are all enclosed in a superconducting solenoidal magnet providing a 1.0~T magnetic field. The solenoid is supported by an octagonal flux-return yoke with resistive plate counter muon identification modules interleaved with steel. 
The charged-particle momentum resolution at $1~{\rm GeV}/c$ is $0.5\%$, and the ${\rm d}E/{\rm d}x$ resolution is $6\%$ for electrons from Bhabha scattering. The EMC measures photon energies with a resolution of $2.5\%$ ($5\%$) at $1$~GeV in the barrel (end cap) region. The time resolution in the plastic scintillator TOF barrel region is 68~ps, while that in the end cap region was 110~ps.  The end cap TOF system was upgraded in 2015 using multigap resistive plate chamber technology, providing a time resolution of 60~ps~\cite{etof1,etof2,etof3}.

The beam pipe is a multilayer structure consisting of a 14.4~$\mu$m gold layer, two beryllium layers with thicknesses of 0.8 and 0.6~mm, and a 0.8~mm oil layer sandwiched between the beryllium layers. In this analysis, the target protons are provided by the hydrogen in the oil layer.

Monte Carlo (MC) simulated data samples produced with a {\sc geant4}-based~\cite{geant4} software package, which includes the geometric description of the \mbox{BESIII} detector and the detector response, are used to determine detection efficiencies and to estimate backgrounds. The simulation models the beam energy spread and initial state radiation in the $e^+e^-$ annihilations with the generator {\sc kkmc}~\cite{ref:kkmc1, ref:kkmc2}. The inclusive MC sample includes both the production of the $J/\psi$ resonance and the continuum processes incorporated in {\sc kkmc}. All particle decays are modeled with {\sc evtgen}~\cite{ref:evtgen1, ref:evtgen2} using branching fractions either taken from the Particle Data Group~\cite{pdg}, when available, or otherwise estimated with {\sc lundcharm}~\cite{ref:lundcharm1, ref:lundcharm2}.
Final state radiation from charged final state particles is incorporated using the {\sc photos} package~\cite{photos2}. 
The signal MC sample of $\jnp$ is generated according to the partial-wave analysis (PWA) results in Ref.~\cite{PWA} to determine the tagging efficiency of antineutron, while signal MC samples of $\bar{n}p \rightarrow K^{+}K^{-}\pi^{+}(\pi^{0})$ are generated with a phase space model to determine detection efficiency. In these simulations, the antineutron momentum is determined by the PWA result of $\jnp$, and the target proton is assumed to be at rest. 

\section{Event Selection}
The signal event selection begins with the tagging of the antineutron from the decay $J/\psi \rightarrow p \pi^{-} \bar{n}$. Charged tracks detected in the multilayer drift chamber~(MDC) are required to be within a polar angle ($\theta$) range of $|\cos\theta|<0.93$, where $\theta$ is defined with respect to the $z$-axis, which is the symmetry axis of the MDC. For charged tracks not originating from $\bar{n}p$ scattering, the distance of closest approach to the interaction point (IP) must be less than 10\,cm along the $z$-axis, and less than 1\,cm in the transverse plane. Particle identification~(PID) for charged tracks combines measurements of the energy deposited in the MDC~(d$E$/d$x$) and the flight time in the time-of-flight system~(TOF) to form likelihoods $\mathcal{L}(h)~(h=p,K,\pi)$ for each hadron $h$ hypothesis. Tracks are identified based on the hypothesis with the highest likelihood. 
To better distinguish particles from $J/\psi$ decay and those from $\bar{n}p$ scattering, a vertex fit is performed on the $p \pi^-$,  with the $\chi^2$ of the fit required to be less than 5. A one-constraint kinematic fit is then applied to the $p \pi^-$, constraining the recoil mass to the nominal antineutron mass~\cite{pdg}. At this stage, no additional $\chi^2$ requirement is applied in the kinematic fit. This avoids introducing a peaking background in the $RM(p\pi^-)$ distribution, which could otherwise bias the extraction of the antineutron yield. 
Here, $RM(p\pi^-)$ is the recoil mass of $p \pi^-$ given by $RM(p\pi^-) = \sqrt{|p_{e^+e^-}-p_p-p_{\pi^-}|^2}$, where $p_{e^+e^-}$, $p_p$ and $p_{\pi^-}$ are the four momenta of the initial $e^+e^-$ system, the proton and the pion before kinematic fit, respectively. 

\begin{figure}[htbp]
    \includegraphics[width=0.35\textwidth]{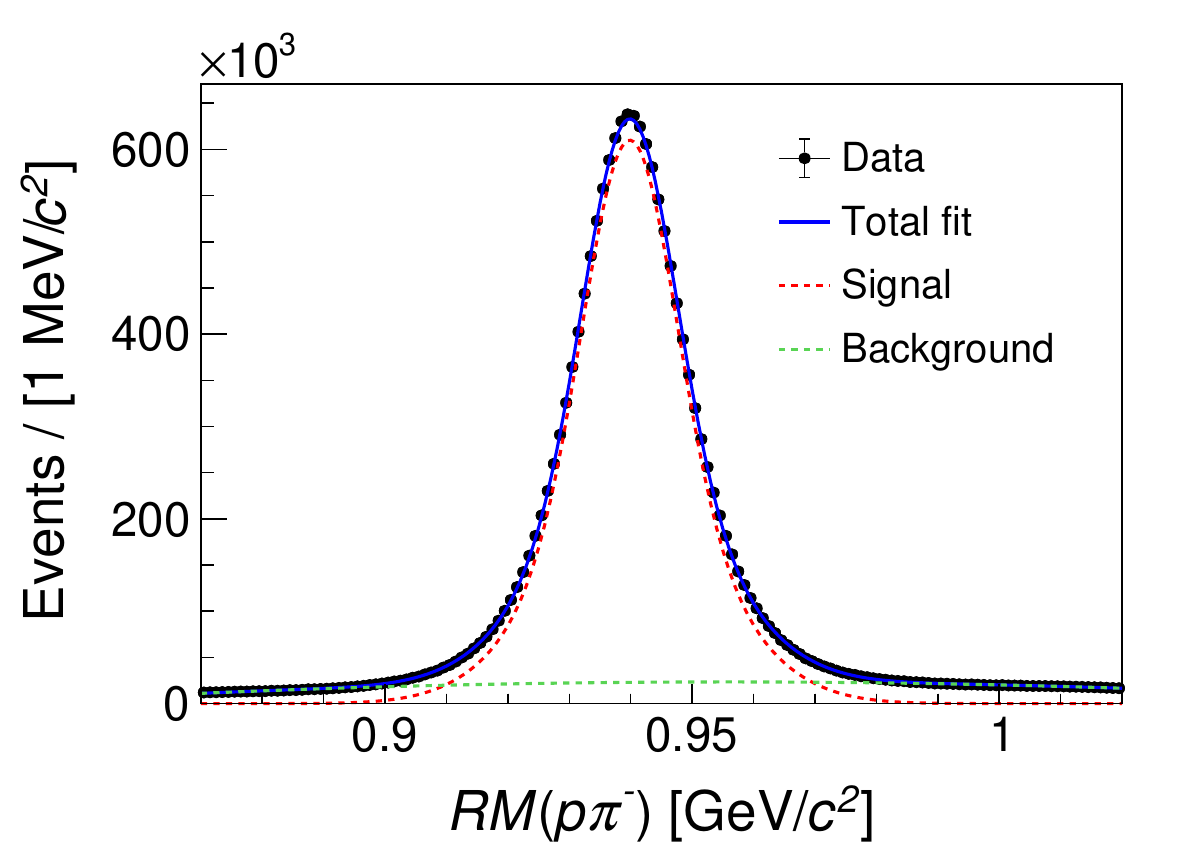}
    \caption{The distribution of $RM(p\pi^-)$. The dots with error bars are data, the blue solid line is the total fit, the red dashed line is the signal shape, and the green dashed line is the background shape.}
    \label{fig:antineutron mass}
\end{figure}

\begin{figure}[htbp]
    \includegraphics[width=0.35\textwidth]{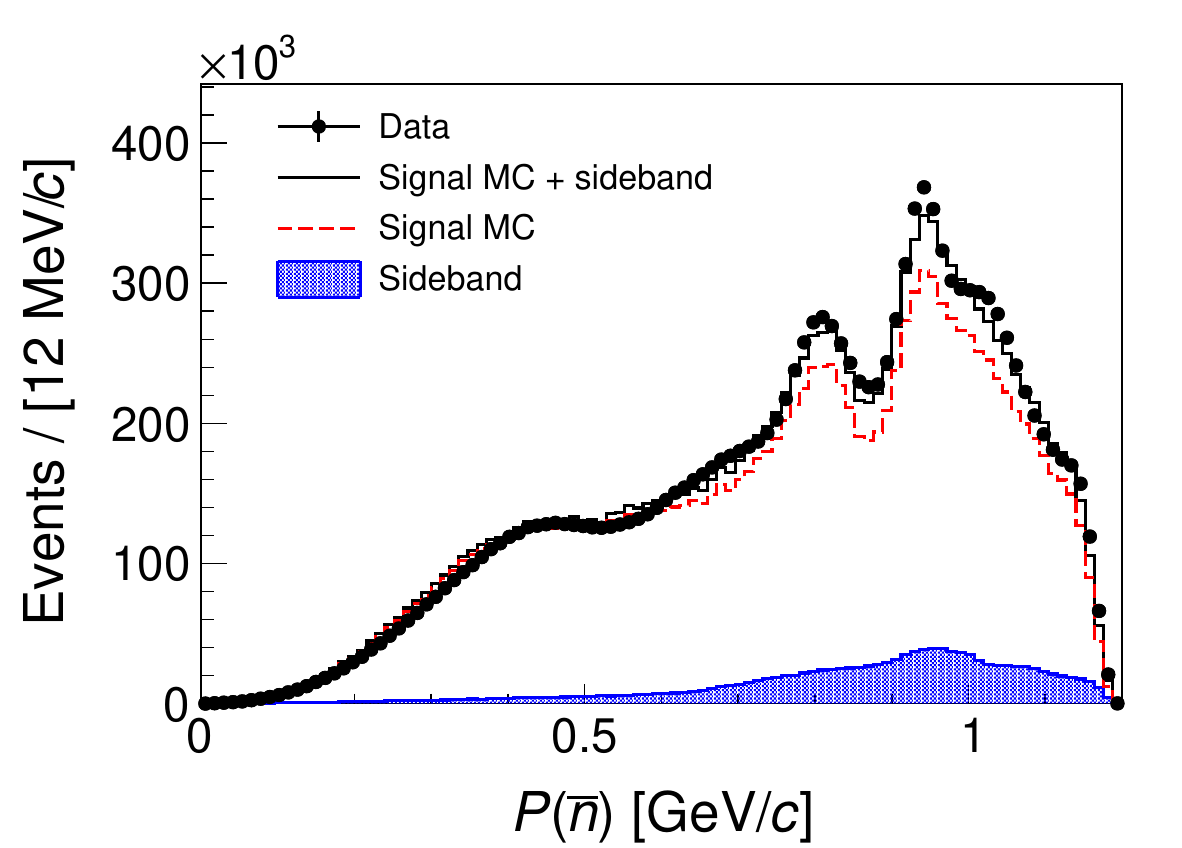}
    \caption{The momentum distribution of tagged antineutrons when $RM(p\pi^-)\in [0.92,0.96]\gevcc$. The sidebands are taken from $RM(p\pi^-)\in [0.88,0.90]\gevcc$ and $[0.98,1.00]\gevcc$, and normalized according to the fit to the $RM(p\pi^-)$ distribution.}
    \label{fig:antineutron momentum}
\end{figure}

To extract the yield of antineutron, a binned maximum likelihood fit is performed on the $RM(p\pi^-)$ distribution with a double-Gaussian function for the signal shape and a second-order Chebyshev polynomial for the background shape. The fit result is shown in Fig.~\ref{fig:antineutron mass}. In the region $RM(p\pi^-)\in [0.89,0.99]\gevcc$, the number of tagged antineutrons and the tagging efficiency are determined to be $N_{\bar{n}} = (14708.9\pm6.4)\times 10^3$ and $\epsilon_{\rm tag} = 66.31\%$, respectively.
The corresponding momentum distribution of the tagged antineutrons, as determined by the kinematic fit, is shown in Fig.~\ref{fig:antineutron momentum}.  
This figure uses a more restricted region, $RM(p\pi^-)\in [0.92,0.96]\gevcc$, for higher purity. The residual background is estimated with sideband events in $RM(p\pi^-)\in [0.88,0.90]\gevcc$ and $[0.98,1.00]\gevcc$, normalized to the expected background contribution in the signal region according to the fit to the $RM(p\pi^-)$ distribution. After adding the normalized sideband contribution, the MC prediction describes the data well, indicating a good consistency between data and the signal MC modeling of the antineutron momentum spectrum.
The distribution is not expected to be smooth, since the $p\pi^-$ system in $\jnp$ contains structures from intermediate baryon resonances, as observed in previous studies~\cite{BES:2004gwe}. This distribution covers a broad momentum range, offering a valuable opportunity to study $\bar{n}p$ scattering for antineutron momentum above $500\mevc$, where existing experimental data are limited.

\begin{figure}[htbp]
    \includegraphics[width=0.23\textwidth]{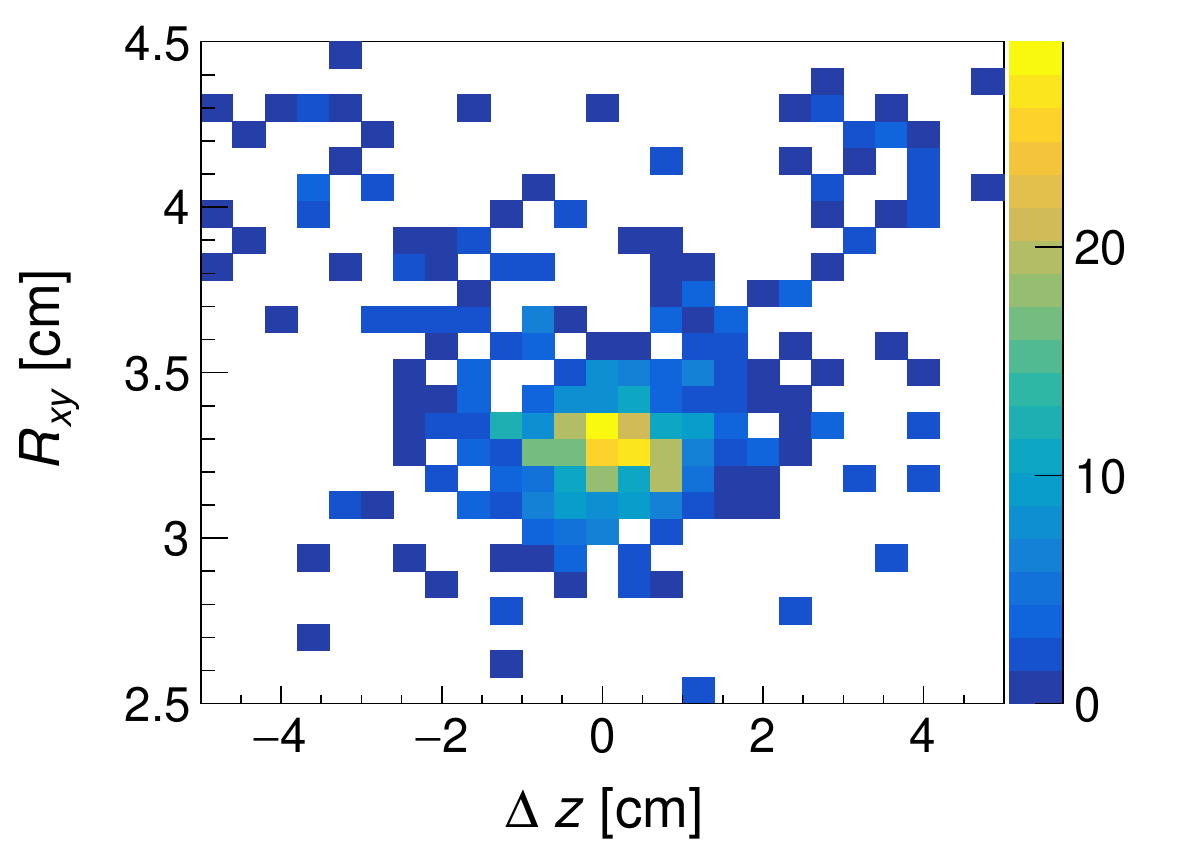}
    \includegraphics[width=0.23\textwidth]{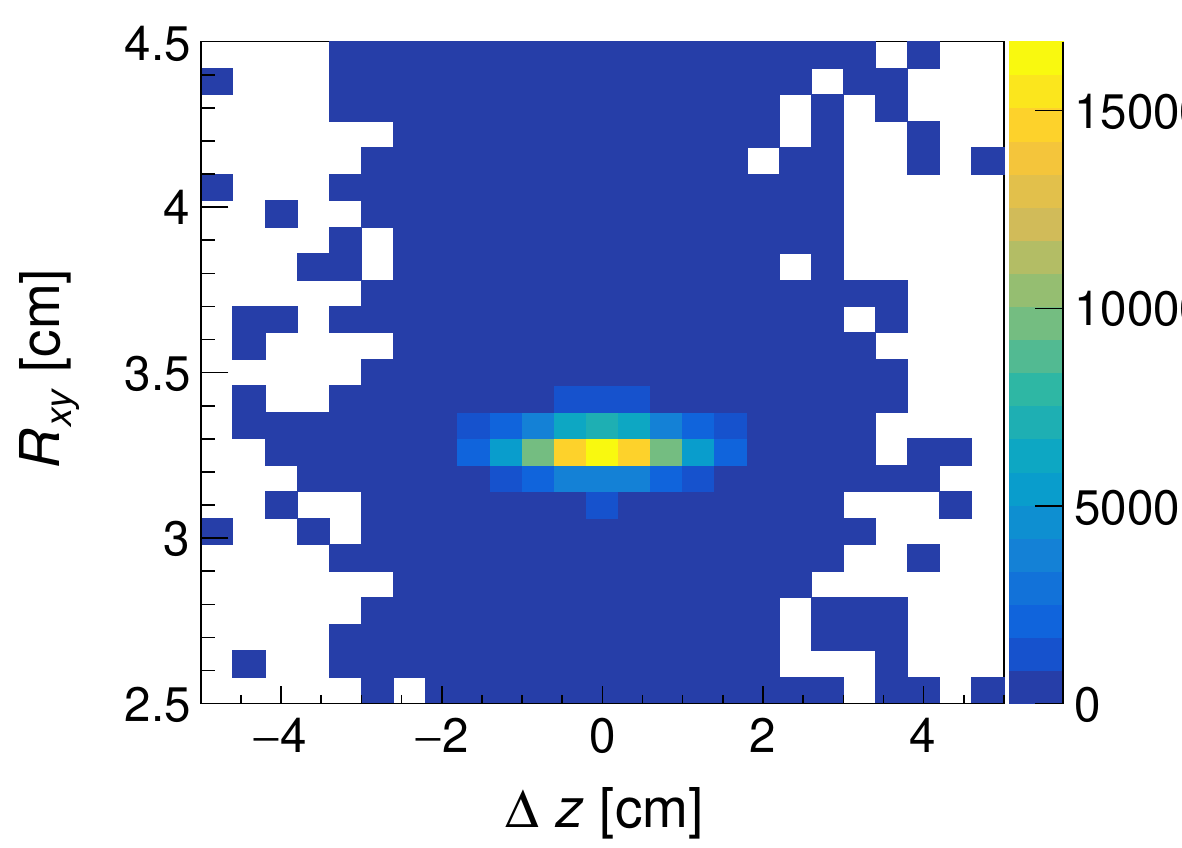}
    \caption{The two-dimensional distributions of the reconstructed transverse radius $R_{xy}$ versus the longitudinal difference $\Delta z$ for data (left) and signal MC (right). In both data and signal MC, an enhanced population is observed near the beam-pipe radius ($R_{xy}\in [3.15,3.37]~\mathrm{cm}$) and around $\Delta z=0$, consistent with antineutron interactions in the beam pipe. The broader $R_{xy}$ distribution in data reflects additional interactions in non-hydrogen beam-pipe materials.}  
    \label{fig:Rxy}
\end{figure}

After tagging the antineutron, we reconstruct $\bar{n}p \rightarrow K^{+}K^{-}\pi^{+}(\pi^{0})$ among the remaining charged tracks and photon candidates. Charged tracks from $\bar{n}p$ scattering are required to be within the polar angle range of $|\cos\theta|<0.93$, with no requirements applied on the distance of closest approach to the IP. After the PID requirement, events are selected if the remaining tracks are consistent with $K^+K^-\pi^+$. Photon candidates are identified using isolated showers in the electromagnetic calorimeter~(EMC). The deposited energy of each shower must be more than 25~MeV in the barrel region ($|\cos \theta|< 0.80$) and more than 50~MeV in the end cap region ($0.86 <|\cos \theta|< 0.92$). To exclude showers that originate from charged tracks, the angle subtended by the EMC shower and the position of the closest charged track at the EMC must be greater than 10 degrees as measured from the IP. To suppress electronic noise and showers unrelated to the event, the difference between the EMC time and the event start time is required to be within [0, 700]\,ns. 
Neutral pion candidates are reconstructed via $\pi^0\rightarrow \gamma \gamma$ with $0.115\gevcc<M(\gamma\gamma)<0.150\gevcc$, and a one-constraint kinematic fit is applied to constrain the invariant mass of the photon pair to the nominal $\pi^0$ mass~\cite{pdg}. If multiple $\pi^0$ candidates are reconstructed, the one with the smallest $\chi^2$ is selected. To suppress non-antineutron background, the $\chi^2$ of the one-constraint kinematic fit on $p \pi^-$ is required to be less than 11. A vertex fit is performed on the $K^+K^-\pi^+$ combination, with its $\chi^2$ required to be less than 6; the resulting vertex position is taken as the $\bar{n}p$ scattering point $(x_1,y_1,z_1)$. Both $\chi^2$ thresholds are optimized based on the figure of merit, $S/\sqrt{S+B}$ for the $\bar{n}p \to K^+K^-\pi^+$ channel, where $S$ is the number of events in the signal MC after all selections and $S+B$ is determined from the inclusive MC sample.
To test whether the reconstructed secondary vertex is consistent with an antineutron interaction in the beam pipe, the fitted vertex position $(x_1, y_1, z_1)$ is compared with the expected antineutron impact point on the beam pipe, $(x_2, y_2, z_2)$, obtained by extrapolating the antineutron flight direction. The reconstructed transverse radius is defined as $R_{xy}=\sqrt{x_1^2+y_1^2}$, and the consistency between the reconstructed and expected positions is quantified by the transverse and longitudinal differences, $\Delta r = R_{xy} - \sqrt{x_2^2+y_2^2}$ and $\Delta z=z_1-z_2$, respectively. The two-dimensional distributions of $R_{xy}$ versus $\Delta z$ shown in Fig.~\ref{fig:Rxy} exhibit a pronounced concentration near the beam-pipe radius and around $\Delta z=0$ in data, consistent with the signal MC expectation. The $R_{xy}$ distribution is broader in data than in signal MC because the data contain antineutron interactions in all beam-pipe materials, including the beryllium layers, whereas the signal MC includes only $\bar{n}p$ interactions on hydrogen nuclei in the oil layer, which constitute the signal process of interest. To suppress combinatorial and non-scattering backgrounds, events are required to satisfy $|\Delta r| < 0.3~\mathrm{cm}$ and $|\Delta z| < 2.0~\mathrm{cm}$.

\begin{figure}[htbp]
    \includegraphics[width=0.23\textwidth]{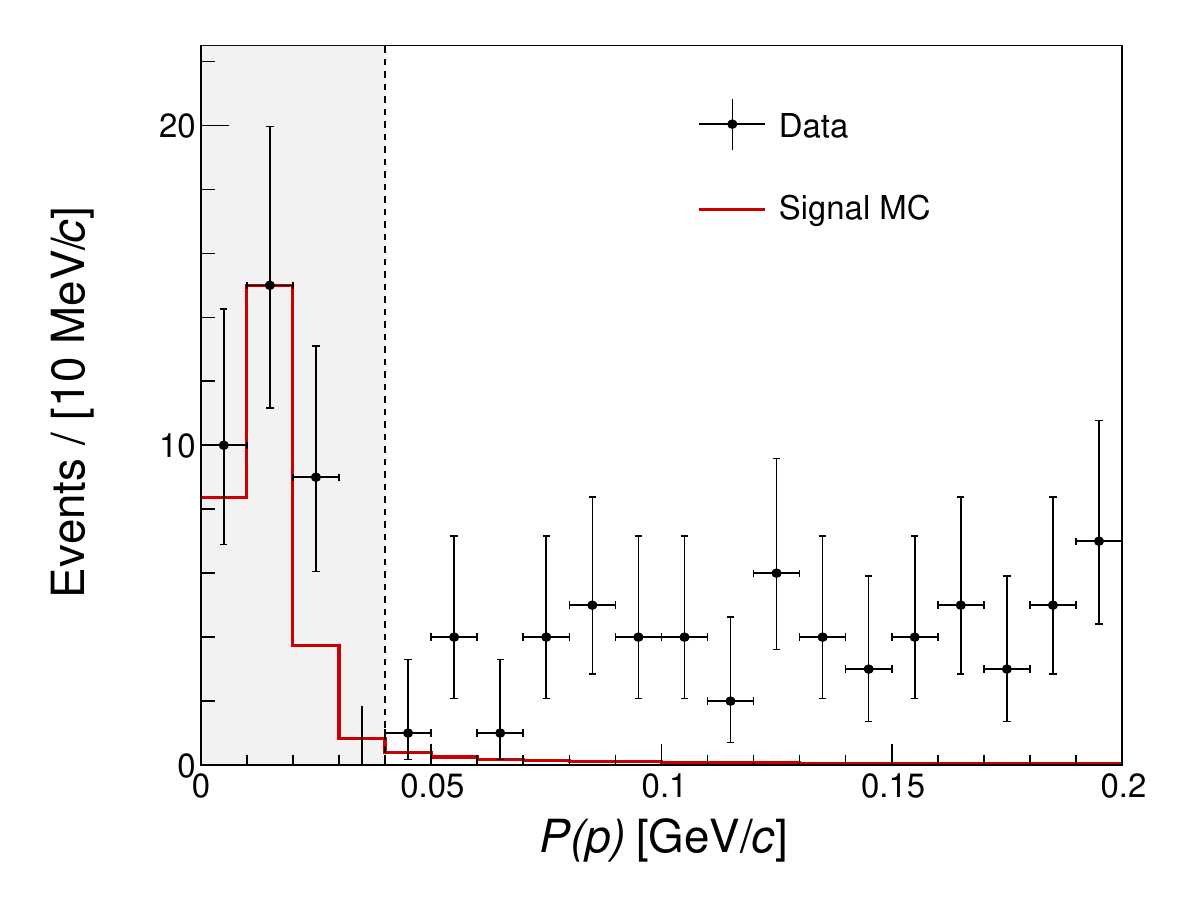}
    \includegraphics[width=0.23\textwidth]{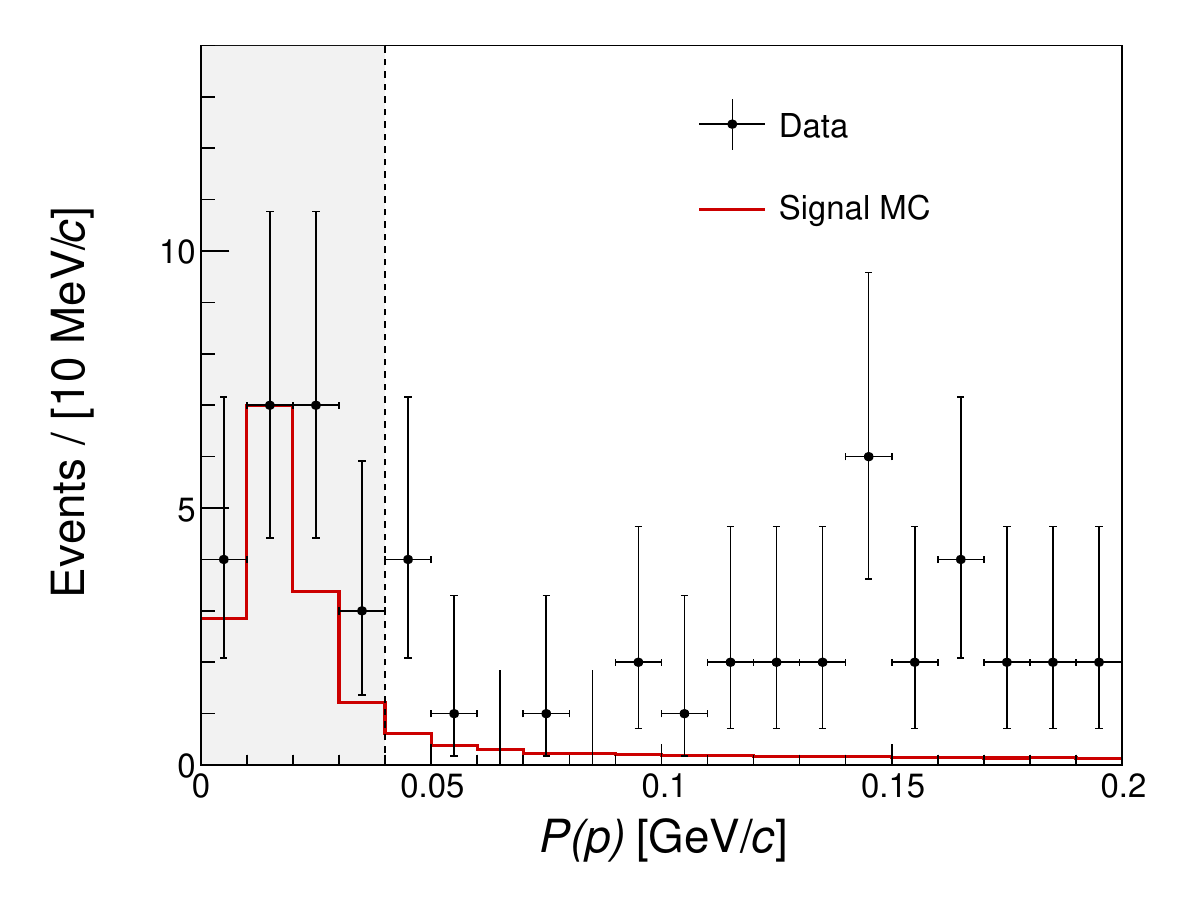}
    \caption{The distributions of $P(p)$ for $\bar{n}p \rightarrow K^{+}K^{-}\pi^{+}$ (left) and $\bar{n}p \rightarrow K^{+}K^{-}\pi^{+}\pi^0$ (right) after the $R_{xy}$ and $\Delta z$ requirements. The dots with error bars are data, and the red histograms are signal MC. The shaded regions indicate the selected low-$P(p)$ intervals, with the vertical dashed lines marking the requirement $P(p)<0.04\gevc$.}
    \label{fig:P_p}
\end{figure}

While the spatial requirements restrict the interactions to the beam pipe, the retained events still contain background contributions from antineutrons scattering with heavier nuclei in the beam-pipe materials. To isolate the signal process of scattering on hydrogen ($^1\mathrm{H}$) in the oil layer, we define the momentum of the target proton as:
\begin{equation}
    P(p) = |\vec{P}(K^+K^-\pi^+(\pi^0)) - \vec{P}(\bar{n})|,
\end{equation}
where $\vec{P}(K^+K^-\pi^+(\pi^0))$  and $\vec{P}(\bar{n})$ are the momenta of the $K^+K^-\pi^+(\pi^0)$ system and the incident antineutron respectively. For events from scattering with protons, $P(p)$ is expected to be close to zero due to the negligible momentum of $^1\rm{H}$ in the oil layer, while for events from scattering with nuclei, $P(p)$ tends to be larger because of the significant Fermi-momentum of nucleons in nuclei. The $P(p)$ distributions after the $R_{xy}$ and $\Delta z$ requirements are shown in Fig.~\ref{fig:P_p} for the $\bar{n}p \rightarrow K^{+}K^{-}\pi^{+}$ and $\bar{n}p \rightarrow K^{+}K^{-}\pi^{+}\pi^0$ channels. The signal MC is concentrated near zero, and the data show an accumulation in the same low-$P(p)$ region together with a broad distribution at larger $P(p)$, which is expected to contain substantial non-hydrogen contributions. Since the low-$P(p)$ region retains most of the signal MC and suppresses these additional contributions, we require $P(p) < 0.04\gevc$ to select $\bar{n}p$ scattering events on $^1\rm{H}$.

\section{Cross Section Measurement}
\begin{figure}[htbp]
    \includegraphics[width=0.23\textwidth]{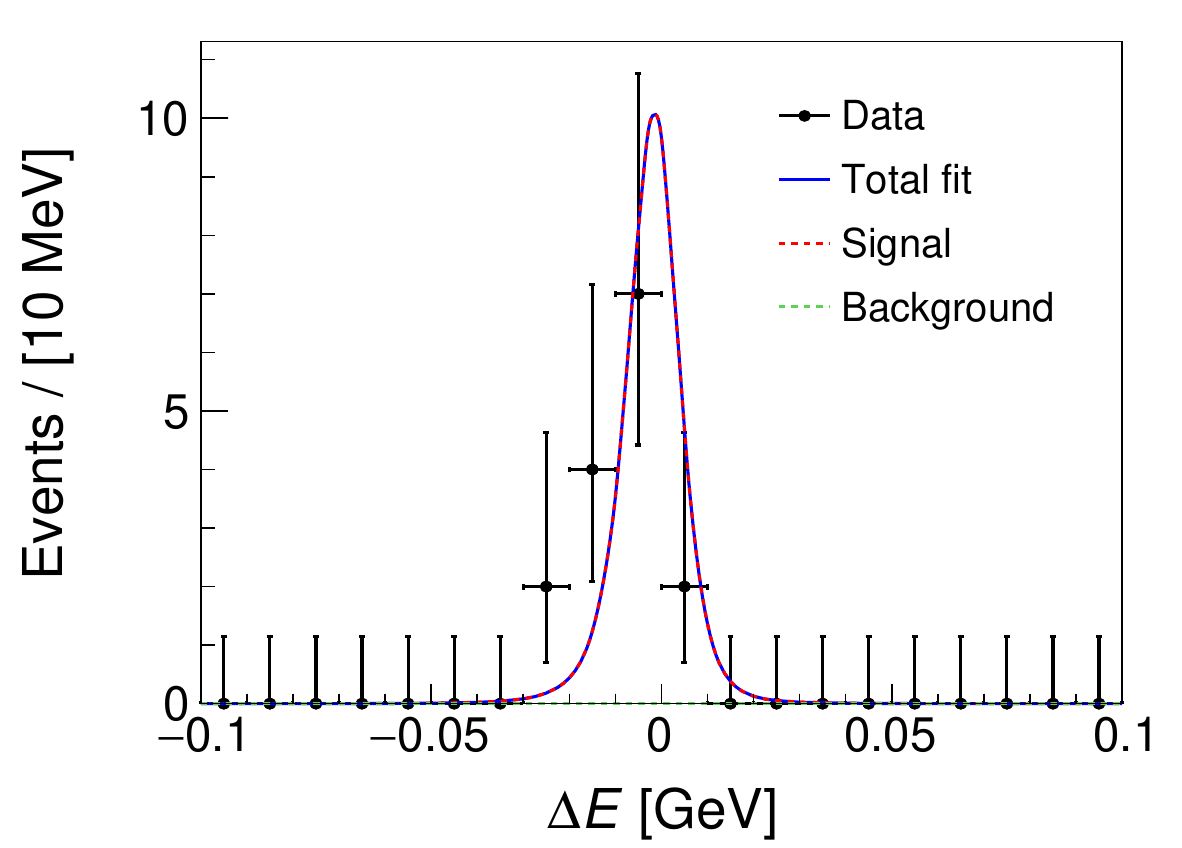}
    \includegraphics[width=0.23\textwidth]{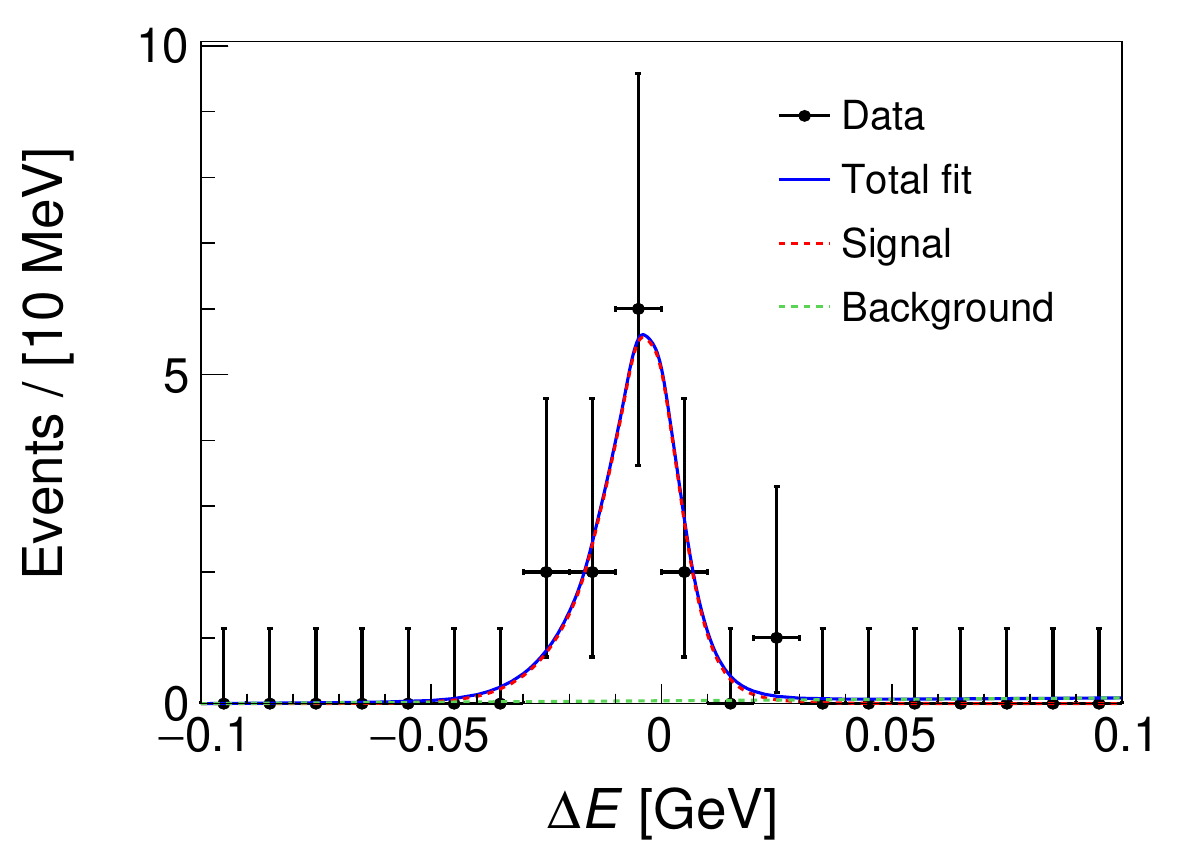}
    \caption{The distributions of $\Delta E$ for $\bar{n}p \rightarrow K^{+}K^{-}\pi^{+}$ (left) and $\bar{n}p \rightarrow K^{+}K^{-}\pi^{+}\pi^0$ (right). The dots with error bars are data, the blue solid lines are the total fits, the red dashed lines are the signal shapes, and the green dashed lines are the background shapes.}
    \label{fig:DeltaE}
\end{figure}

Energy conservation in $\bar{n}p$ scattering requires that the total energy of the final states $E(K^+K^-\pi^+(\pi^0))$ equals the sum of the initial antineutron energy $E_{\bar{n}}$ and the proton mass $m_p$. To distinguish signal from background, we introduce the kinematic variable $\Delta E$ defined as:
\begin{equation}
    \Delta E = E(K^+K^-\pi^+(\pi^0)) - E_{\bar{n}} - m_p .
\end{equation}
An unbinned maximum likelihood fit is performed on the $\Delta E$ distribution to extract the signal yield. In the fit, the signal shape is described by the signal MC shape, while the background shape is described by a first-order Chebyshev polynomial. The distributions of $\Delta E$ and the fit results are shown in Fig.~\ref{fig:DeltaE}. Clear samples of $\bar{n}p$ scattering are observed, and the signal yields are determined to be $15.0^{+4.2}_{-3.5}$ and $12.1^{+4.0}_{-3.3}$ for $\bar{n}p \rightarrow K^{+}K^{-}\pi^{+}$ and $\bar{n}p \rightarrow K^{+}K^{-}\pi^{+}\pi^0$, respectively.

The integrated luminosity of the antineutron beam is calculated as~\cite{CLAS:2021gur}:
\begin{equation}
    \mathcal{L}_{\bar{n}} = \frac{ N_A \, \rho_H \, l_H}{M_H} \, \frac{N_{\bar{n}}}{\epsilon_{\rm tag}} ,
\end{equation}
where $N_A$ is the Avogadro constant, $\rho_H = 0.1221~\rm{g/cm^3}$ is the density of $^1\rm{H}$ in the oil layer, $l_H$ is the average path length of antineutron beam in the oil layer and $M_H$ is the molar mass of $^1\rm{H}$. The quantities $N_{\bar{n}}$ and $\epsilon_{\rm tag}$ are the number of tagged antineutrons and tagging efficiency, respectively. The average path length $l_H$, which depends on the antineutron $\cos \theta$ distribution, is determined to be $l_H = 0.106\cm$ utilizing the PWA result of $\jnp$ in Ref.~\cite{PWA}. The calculated integrated luminosity is $\mathcal{L}_{\bar{n}} = 1.73 \times 10^{29}\ \rm{cm^{-2}}$. The cross section is then given by:
\begin{equation}
    \sigma(\bar{n}p \rightarrow K^{+}K^{-}\pi^{+}(\pi^{0})) = \frac{N_{\rm sig}}{\epsilon \, \mathcal{L}_{\bar{n}} \, \mathcal{B}(\pi^0 \rightarrow \gamma \gamma)} ,
\end{equation}
where $N_{\rm sig}$ is the signal yield obtained from the fit, $\epsilon$ is the selection efficiency, and $\mathcal{B}(\pi^0 \rightarrow \gamma \gamma)$ is the branching fraction of $\pi^0 \rightarrow \gamma \gamma$~\cite{pdg}, which is only considered for $\bar{n}p \rightarrow K^{+}K^{-}\pi^{+}\pi^0$. The simulation is re-weighted according to the $\cos\theta$ and momentum distributions of antineutrons observed in data to determine the efficiencies. The resulting efficiencies are $\epsilon = 16.40\%$ and $6.52\%$ for $\bar{n}p \rightarrow K^{+}K^{-}\pi^{+}$ and $\bar{n}p \rightarrow K^{+}K^{-}\pi^{+}\pi^0$, respectively. Based on these, the flux-averaged cross sections over the antineutron momentum spectrum are determined to be $\sigma(\bar{n}p \rightarrow K^{+}K^{-}\pi^{+}) = 0.53^{+0.15}_{-0.12}$ mb and $\sigma(\bar{n}p \rightarrow K^{+}K^{-}\pi^{+}\pi^0) = 1.09^{+0.36}_{-0.30}$ mb, where the uncertainties are statistical only.

\begin{figure}[htbp]
    \includegraphics[width=0.23\textwidth]{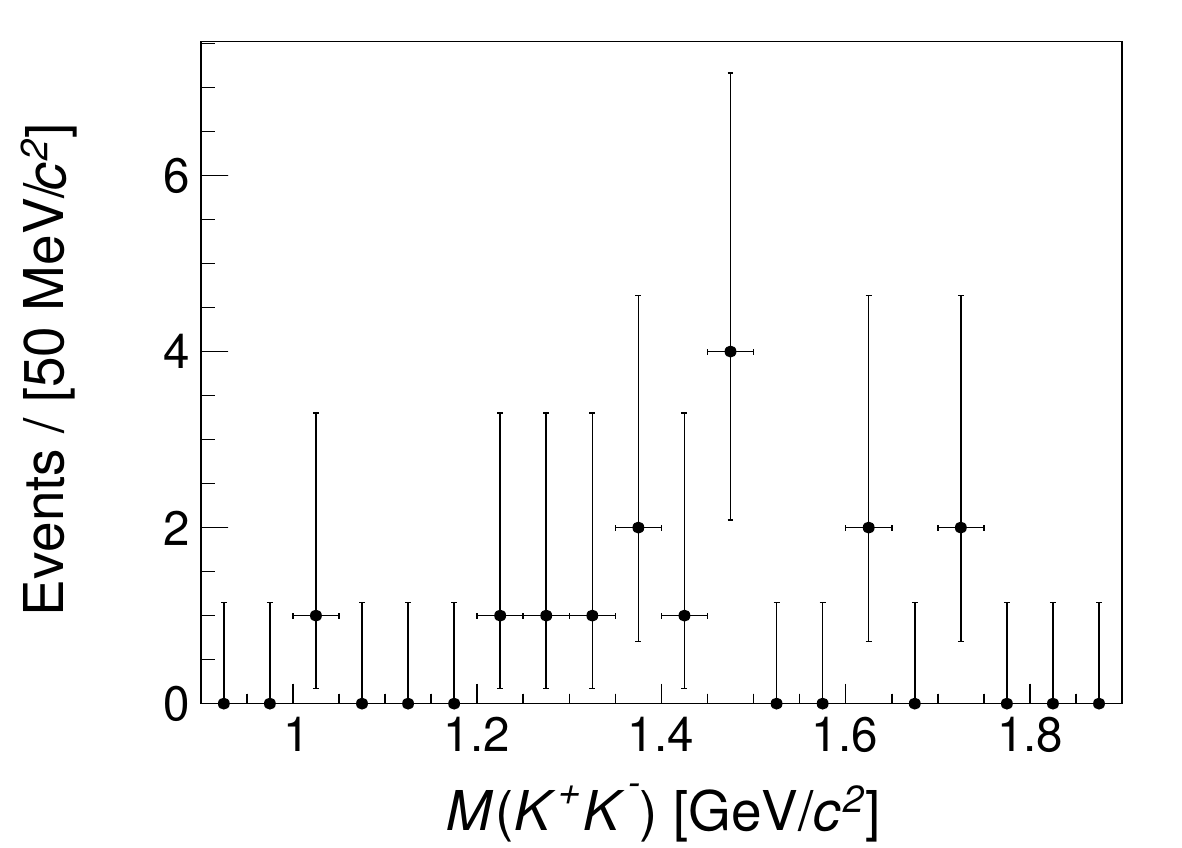}
    \includegraphics[width=0.23\textwidth]{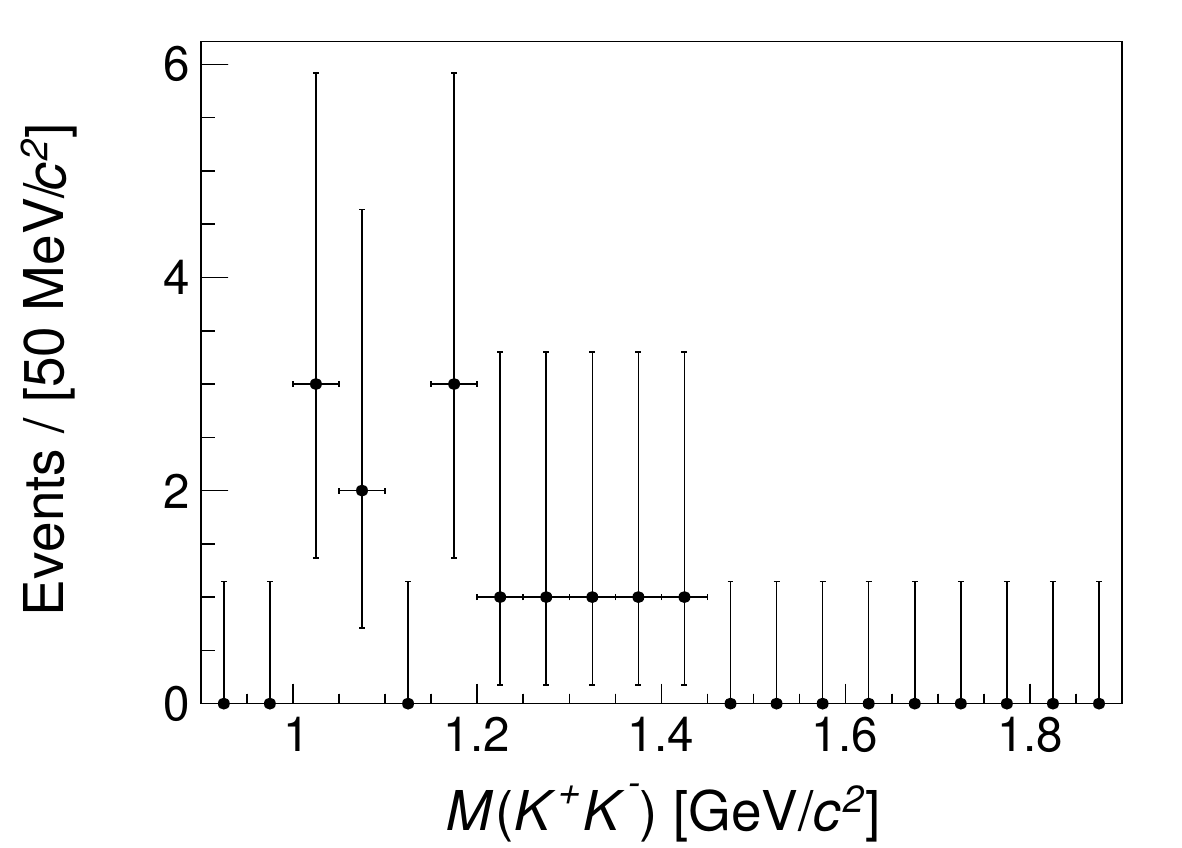}
    \caption{The distributions of the invariant masses of $K^+K^-$ for $\bar{n}p \rightarrow K^{+}K^{-}\pi^{+}$ (left) and $\bar{n}p \rightarrow K^{+}K^{-}\pi^{+}\pi^0$ (right) in data.}
    \label{fig:M_KK}
\end{figure}

A qualitative inspection of possible intermediate states in $\bar{n}p$ scattering is also performed. The distributions of the invariant mass of $K^+K^-$ are shown in Fig.~\ref{fig:M_KK}. The enhancement around 1.5$\gevcc$ and 1.0$\gevcc$ in the $K^+K^-$ invariant mass in $\bar{n}p \rightarrow K^{+}K^{-}\pi^{+}$ and in $\bar{n}p \rightarrow K^{+}K^{-}\pi^{+}\pi^0$ may originate from $f_0(1500)$ and $\phi$, respectively. Due to limited statistics, it is difficult to quantitatively extract the contributions of intermediate states.

\section{Systematic Uncertainties}
The systematic uncertainties in the cross section measurements are evaluated as follows. Uncertainties associated with tracking and PID of the $p\pi^-$ on the tag side, as well as the vertex fitting for the $p\pi^-$, are not considered separately since the tagging efficiency of the antineutron cancels between
the neutron flux and the event yield in the cross section calculation. The uncertainties of tracking and PID in the signal side are estimated to be 1.0\% per track~\cite{trk uncertainty}. The uncertainty from $\pi^{0}$ reconstruction is studied with a control sample of $D^{0} \to K^-\pi^+\pi^0$ decays and is determined to be 2.5\% for $\bar{n}p \to K^+K^-\pi^+\pi^0$.
The uncertainty from the fit to $RM(p\pi^-)$ is estimated by varying the fit range, signal shape and background shape, where the fit range is changed from [0.87, 1.02]$\gevcc$ to [0.85, 1.04]$\gevcc$, the signal shape is changed from double-Gaussian function to the signal MC shape convolved with a Gaussian function, and the background shape is varied from a second-order to a third-order Chebyshev polynomial. The resulting shifts from the nominal fit are then summed in quadrature.  To estimate the uncertainties from the requirements on the $\chi^2$ of kinematic fit and vertex fit for $K^+K^-\pi^+$, scattering position and $P(p)$,  the selection criteria are varied by $\pm10\%$, and the largest resulting changes in cross sections are taken as the systematic uncertainties. 
The uncertainty from signal yield extraction is estimated by varying the signal MC shape to the signal MC distribution convolved with a Gaussian function in the $\Delta E$ fit. 
Despite the $P(p)$ selection, a small fraction of background events from $\bar{n}$ scattering on other nuclei ($^{9}\mathrm{Be}$ and $^{12}\mathrm{C}$) may persist. To estimate the systematic uncertainty associated with this residual background, we evaluate the event yield in a control region of $0.04 \le P(p) < 0.30~\gevc$ after performing a $\Delta E$ sideband subtraction. This yield is subsequently extrapolated to the signal region ($P(p) < 0.04~\gevc$) utilizing the nucleon momentum distributions within the nuclei, as detailed in Ref.~\cite{Fermi momentum}. The resulting changes in the cross sections are taken as the systematic uncertainties.
Intermediate states are not considered in the generation of signal MC samples. The uncertainty associated with such states is estimated by re-weighting the efficiency according to the momentum and $\cos\theta$ for all final-state particles, and calculated as $\sigma = \sqrt{\sum_i \sigma_i^2}$, where $\sigma_i$ is the change in cross section caused by one final state momentum or $\cos \theta$ weighting. 
A summary of all systematic uncertainties is shown in Table~\ref{tab:sys}. The total systematic uncertainties are determined to be 14.8\% and 28.4\% for $\bar{n}p \rightarrow K^{+}K^{-}\pi^{+}$ and $\bar{n}p \rightarrow K^{+}K^{-}\pi^{+}\pi^0$, respectively, obtained by adding all individual contributions in quadrature.

\begin{table}[!htb] 
    \caption{The systematic uncertainties in the cross section measurements  (in \%). The symbol ``$-$'' indicates not applicable. $K^{+}K^{-}\pi^{+}$ and $K^{+}K^{-}\pi^{+}\pi^0$ represent $\bar{n}p \rightarrow K^{+}K^{-}\pi^{+}$ and $\bar{n}p \rightarrow K^{+}K^{-}\pi^{+}\pi^0$, respectively.}
    \centering
    \begin{tabular}{c|cc}
    \hline \hline
        Source                                      &$K^{+}K^{-}\pi^{+}$  &$K^{+}K^{-}\pi^{+}\pi^0$    \\ \hline 
        Tracking                                    & 3.0                 &  3.0                \\ 
        PID                                         & 3.0                 &  3.0                \\
        $\pi^0$ reconstruction                      & -                   &  2.5                \\
        $RM(p\pi^-)$ fit                            & 1.6                 &  1.6                \\
        $\chi^2$ cut on kinematic fit               & 6.0                 &  0.7                \\   
        $\chi^2$ cut on vertex fit of $p\pi^-$      & 1.6                 &  0.8                \\  
        Scattering position cut                     & 1.8                 &  2.1                \\  
        $P(p)$ requirement                          & 0.8                 &  7.3                \\  
        Signal yield extraction                     & 0.1                 &  7.4                \\  
        Background from other nuclei                & 2.7                 &  1.7                \\
        Intermediate states                         & 12.3                &  25.7               \\ \hline  
        Total                                       & 14.8                &  28.4               \\ \hline \hline
    \end{tabular}
    \label{tab:sys}
\end{table}

\section{Summary}
In summary, we report measurements of the flux-averaged cross sections for the reactions $\bar{n}p \rightarrow K^{+}K^{-}\pi^{+}$ and $\bar{n}p \rightarrow K^{+}K^{-}\pi^{+}\pi^0$ using a novel antineutron-production method at the BESIII detector. The antineutrons are produced via $J/\psi \rightarrow p \pi^{-} \bar{n}$, with momenta ranging from 0 to 1174$\mevc$, and the target protons are provided by the hydrogen in the beam-pipe cooling oil. The measured cross sections are $0.53^{+0.15}_{-0.12} \pm 0.08$~mb and $1.09^{+0.36}_{-0.30} \pm 0.31$~mb, respectively, where the first uncertainties are statistical and the second are systematic. For comparison, the isospin-related process $p\bar{p} \rightarrow K^+K^-\pi^0$ was measured to have a cross section of $0.347 \pm 0.037$~mb at $900\mevc$~\cite{CrystalBarrel:2006mhj}. The corresponding partial-wave analysis indicates that the $K\bar{K}$ resonance contribution at this momentum is dominated by the isoscalar component. Under this assumption, isospin symmetry gives $\sigma(\bar{n}p \rightarrow K^{+}K^{-}\pi^{+}) = 2\sigma(p\bar{p} \rightarrow K^+K^-\pi^0)$, this expectation is consistent with the present measurement within uncertainties, although the comparison is qualitative because the present result is averaged over a broad antineutron momentum spectrum and the limited statistics do not allow differential measurements in momentum bins.
Although limited statistics prevent a detailed study of intermediate states, the observed $\bar{n}p$ scattering candidates demonstrate the feasibility of using antineutrons from $J/\psi$ decays to investigate strangeness production in antineutron-proton interactions. Proposed future $e^+e^-$ colliders in the tau-charm energy region, such as STCF and SCTF~\cite{stcf,sctf}, could significantly improve the quality and yield of antineutron sources, and allow for optimization of target materials, enabling more precise measurements of antineutron physics.

\begin{acknowledgments}
The BESIII Collaboration thanks the staff of BEPCII (https://cstr.cn/31109.02.BEPC) and the IHEP computing center for their strong support. This work is supported in part by National Key R\&D Program of China under Contracts Nos. 2023YFA1606000, 2023YFA1606704; National Natural Science Foundation of China (NSFC) under Contracts Nos. 11635010, 11935015, 11935016, 11935018, 12025502, 12035009, 12035013, 12061131003, 12192260, 12192261, 12192262, 12192263, 12192264, 12192265, 12221005, 12225509, 12235017, 12342502, 12361141819; the Chinese Academy of Sciences (CAS) Large-Scale Scientific Facility Program; the Strategic Priority Research Program of Chinese Academy of Sciences under Contract No. XDA0480600; CAS under Contract No. YSBR-101; 100 Talents Program of CAS; The Institute of Nuclear and Particle Physics (INPAC) and Shanghai Key Laboratory for Particle Physics and Cosmology; ERC under Contract No. 758462; German Research Foundation DFG under Contract No. FOR5327; Istituto Nazionale di Fisica Nucleare, Italy; Knut and Alice Wallenberg Foundation under Contracts Nos. 2021.0174, 2021.0299, 2023.0315; Ministry of Development of Turkey under Contract No. DPT2006K-120470; National Research Foundation of Korea under Contract No. NRF-2022R1A2C1092335; National Science and Technology fund of Mongolia; Polish National Science Centre under Contract No. 2024/53/B/ST2/00975; STFC (United Kingdom); Swedish Research Council under Contract No. 2019.04595; U. S. Department of Energy under Contract No. DE-FG02-05ER41374.
\end{acknowledgments}

\end{document}